\documentclass[aps,prd,twocolumn,superscriptaddress,amsmath,amssymb,amsthm,nofootinbib, preprintnumbers]{revtex4-2}
\usepackage{url}
\usepackage[colorlinks=true,
    linkcolor=blue,
    citecolor=blue,
    urlcolor=blue]{hyperref}
 \usepackage{amsmath}
\usepackage{graphicx}
\usepackage{epstopdf}
\usepackage{float}
\usepackage{physics}
\usepackage{color}
\usepackage[T1]{fontenc}
\usepackage[utf8]{inputenc}
\usepackage[toc,page]{appendix}
\usepackage[usenames,dvipsnames]{xcolor}
\usepackage[normalem]{ulem}
\usepackage{hyperref}

\usepackage{tikz}
\usetikzlibrary{shapes,positioning,shadows,graphs.standard,automata,arrows}
\usepackage{pgfplots}
\pgfplotsset{compat=newest, width=2.669cm, height=2.669cm, scale only axis=true,enlargelimits=false}
\pgfplotsset{tick label style={font=\tiny}}
\pgfplotsset{every major tick/.append style={major tick length=3pt}}
\pgfplotsset{every minor tick/.append style={minor tick length=1.5pt}}
\usepgfplotslibrary{groupplots}
\usepackage{caption}
\usepackage[font=small,labelfont=bf,justification=justified,format=plain]{caption}

\providecommand{\renewoperator}[3]{%
\renewcommand*{#1}{\mathop{#2}#3}}

\DeclareMathOperator{\artanh}{artanh}

\makeatletter
\providecommand*{\diff}%
{\@ifnextchar^{\DIfF}{\DIfF^{}}}
\def\DIfF^#1{%
\mathop{\mathrm{\mathstrut d}}%
\nolimits^{#1}\gobblespace}
\def\gobblespace{%
\futurelet\diffarg\opspace}
\def\opspace{%
\let\DiffSpace\!%
\ifx\diffarg(%
\let\DiffSpace\relax
\else
\ifx\diffarg[%
\let\DiffSpace\relax
\else
\ifx\diffarg\{%
\let\DiffSpace\relax
\fi\fi\fi\DiffSpace}

\renewoperator{\Re}{\mathrm{Re}}{\nolimits}
\renewoperator{\Im}{\mathrm{Im}}{\nolimits}

\usepackage{bm}

\newcommand{\be}{\begin{equation}}
\newcommand{\ee}{\end{equation}}
\newcommand{\ba}{\begin{eqnarray}}
\newcommand{\ea}{\end{eqnarray}}

\newcommand{\beq}{\begin{equation}}
\newcommand{\eeq}{\end{equation}}
\newcommand{\beqa}{\begin{eqnarray}}
\newcommand{\eeqa}{\end{eqnarray}}

\begin{document}

\title{A Circular Disformal Kerr Black Hole}  

\author{Jibril Ben Achour}
\email{j.benachour@lmu.de}
\affiliation{Arnold Sommerfeld Center for Theoretical Physics, Munich, Germany}
\affiliation{Univ de Lyon, ENS de Lyon, Laboratoire de Physique, CNRS UMR 5672, Lyon 69007, France}

\author{Adolfo Cisterna}
\email{adolfo.cisterna.r@mail.pucv.cl}
\affiliation{Sede Esmeralda, Universidad de Tarapac{\'a}, Avenida Luis Emilio Recabarren 2477, Iquique, Chile}

\author{Hugo Roussille}
\email{hugo.roussille@ens-lyon.fr}
\affiliation{Univ de Lyon, ENS de Lyon, Laboratoire de Physique, CNRS UMR 5672, Lyon 69007, France}

\date{\today} 
 
\begin{abstract}

The Kerr solution is the cornerstone of General Relativity (GR) for modeling astrophysical rotating black holes and for testing GR through gravitational-wave observations and black hole imaging. Understanding how the Kerr geometry is modified in alternative theories of gravity is therefore a crucial step toward constraining possible deviations from GR. Despite their importance, exact analytical solutions describing rotating black holes in modified gravity are rare, limiting our ability to explore novel phenomenology and to design robust observational tests of new physics.
In this work, we present a new exact rotating black hole solution within a specific scalar–tensor theory belonging to the Horndeski class. The solution is obtained via a disformal transformation acting on a Kerr stealth black hole. Crucially, unlike previous constructions, the disformal transformation of our chosen seed configuration preserves circularity, ensuring that many of the geometrical and physical properties that make the Kerr spacetime so compelling are retained. We refer to the resulting geometry as the Circular Disformal Kerr solution.
Remarkably, key features such as the structure of Killing horizons, the ergosphere, and the absence of causality violations closely mirror those of the Kerr metric. The spacetime is algebraically general, corresponding to Petrov type I. This new exact solution therefore provides a rare example of a rotating black hole beyond GR that closely mimics the Kerr geometry, offering a valuable theoretical laboratory to investigate the phenomenology of Kerr-like black holes in modified gravity.

\end{abstract}

\maketitle  

\makeatletter
\makeatother

\newpage

\section*{Introduction}

Parallel advances in gravitational-wave astronomy and black hole imaging have opened an unprecedented window for testing our theoretical models of compact objects \cite{LIGOScientific:2020tif, EventHorizonTelescope:2019dse,EventHorizonTelescope:2019ggy, Johnson:2024ttr}. To date, these observations have confirmed the predictions of General Relativity (GR), indicating that the spacetime geometry of astrophysical black holes is well described by the Kerr solution. However, fully exploiting the constraining power of current and future observations to test possible deviations from GR requires a precise characterization of the phenomenology of rotating black holes beyond it.
A central question in this context is to determine to what extent present and forthcoming observations can distinguish the Kerr solution of GR from alternative Kerr mimickers that may reproduce similar observational signatures. A standard strategy to address this issue consists of introducing \textit{ad hoc} deviations in the metric through free functions, which are subsequently constrained by observational data or to consider the presence of torsion in the spacetime \cite{Johannsen:2011dh, Johannsen:2013szh,Johannsen:2013vgc, Cardoso:2014rha,Papadopoulos:2018nvd, Carson:2020dez, BenAchour:2025uzp,Ghosh:2024arw,Heinicke:2014ipp}. While this approach can be justified as a first exploratory step, it suffers from several limitations. First, the deviations introduced in such \textit{ad hoc} parametrizations generally lack a clear geometrical interpretation. Second, they only provide tests of a given family of geometries and, hence, do not provide a direct test of modified theories of gravity.
To genuinely test extensions of GR, a complementary but considerably more challenging approach consists of constructing exact analytical solutions of modified gravity theories that describe rotating black holes and extracting their physical and observational properties. To fully appreciate the difficulty of this task, it is instructive to review the status of the Kerr solution and discuss the various obstacles encountered when searching for Kerr-like solutions beyond GR.

The Kerr black hole \cite{Kerr:1963ud} represents the unique stationary end state of a sufficiently massive, rotating star under the assumptions of asymptotic flatness and a regular, non-degenerate event horizon \cite{Carter:1971zc}. This uniqueness is encoded in the famous no-hair theorem. Within GR, two notable examples of rotating vacuum solutions beyond the Kerr geometry are known, the Tomimatsu–Sato (TS) \cite{Tomimatsu:1972zz} and Kerr–Levi-Civita (Kerr-LC) \cite{Barrientos:2025rjn} classes, each possessing a well-defined static limit. The TS family, while asymptotically flat, evades the uniqueness theorem by introducing a non-trivial multipolar extension of Kerr, which leads to naked singularities where the event horizon would normally form \cite{Kodama:2003ch}. Its static limit is the Zipoy–Voorhees spacetime \cite{Zipoy:1966btu,Voorhees:1970ywo}, where the multipolar structure is more pronounced \cite{Quevedo:2012ttw}.
The Kerr-LC family, in contrast, is a fully regular solution embedding the rotating source in an asymptotically Levi-Civita background \cite{Stephani:2003tm}. It provides a regular, rotating black hole in vacuum, with its static limit given by the Schwarzschild–Levi-Civita spacetime~\cite{Akbar:2015qna,Mazharimousavi:2024hrg,Barrientos:2024uuq}.

Rotating solutions are notoriously difficult to obtain analytically. The discovery of Kerr relied on the Goldberg–Sachs theorem \cite{GStheorem}, which characterizes algebraically special solutions via a geodesic, shear-free principal null direction, and on the Robinson–Trautman family of expanding, shear- and twist-free spacetimes \cite{Robinson:1962zz}. Similarly, constructing TS and Kerr-LC solutions requires specialized techniques, most notably the Ernst formalism for electrovacuum configurations \cite{Ernst:1967by,Ernst:1967wx} and the associated Kinnersley group of Lie point symmetries \cite{Kinners,Kramer,Geroch:1970nt}.

The inclusion of matter fields further complicates the construction of rotating black hole solutions. Beyond the electrovacuum case --- where certain Lie point symmetries can still be used to describe stationary, axisymmetric black holes in electromagnetic fields \cite{Harrison} --- most matter sources are constrained by no-hair theorems, which enforce trivial field profiles when attempting to “dress” a black hole \cite{Herdeiro:2015waa}. As a result, exact solutions with additional degrees of freedom arising from non-electromagnetic energy–momentum tensors are rare, even for spherically symmetric geometries, and become markedly more challenging once rotation is introduced.

No-hair theorems are inherently theory-dependent, but their consequences often limit both the physical relevance of exact solutions and the range of possible phenomenological applications. For instance, in Einstein gravity coupled to a free, minimally coupled scalar field, the backreaction of the scalar decouples from the Einstein–scalar equations. This allows any stationary, axisymmetric vacuum solution to be endowed with a non-trivial scalar profile \cite{Eris:1976xj}, affecting only the non-Killing sector of the metric while leaving the remaining structure formally identical to the vacuum solution \cite{Barrientos:2025abs}. However, due to the constraints imposed by no-hair theorems \cite{Bekenstein:1995un}, such solutions inevitably exhibit naked singularities, either at the would-be event horizon or at infinity.

With the development of generalized scalar–tensor theories \cite{Horndeski:1974wa,Gleyzes:2014dya,Gleyzes:2014qga,Langlois:2015cwa,Langlois:2015skt,Crisostomi:2016czh,BenAchour:2016fzp}, there has been renewed interest in obtaining exact, well-behaved black hole solutions endowed with scalar hair. This has prompted the formulation of progressively refined no-hair theorems covering increasingly broad classes of models \cite{Hui:2012qt,Maselli:2015yva,Creminelli:2020lxn,Capuano:2023yyh}.
Within this context, an intriguing mechanism has been identified: under certain conditions, the scalar field’s energy–momentum tensor can vanish despite a non-trivial configuration, forcing the spacetime metric to coincide with a vacuum GR solution. These stealth solutions \cite{Ayon-Beato:2004nzi} describe vacuum geometries that support non-trivial scalar fields without backreacting on the metric.

Although stealth configurations may initially seem of limited phenomenological interest, they are important for two reasons. First, their existence depends primarily on specific conditions of the underlying theory rather than on the spacetime geometry itself \cite{BenAchour:2018dap,Motohashi:2019sen,Takahashi:2020hso,Charmousis:2019vnf}. Second, they provide well-behaved seed solutions for frame transformations or mappings between scalar–tensor theories, enabling the construction of genuinely backreacting configurations with novel geometric features \cite{BenAchour:2020wiw}. 
It should be noted that, in scalar–tensor theories, the standard solution-generating techniques developed for stationary and axisymmetric electrovacuum spacetimes do not directly apply. Frame transformations and related mappings, therefore, serve as alternative tools to explore and construct non-trivial solutions in these extended gravitational frameworks \cite{BenAchour:2020fgy, Anson:2020trg}.

The application of a disformal transformation does not require the seed solution to be of the stealth type; it can also act on configurations with nontrivial backreaction. However, most such non-stealth seeds typically exhibit pathologies (often curvature singularities) that are generally preserved under the transformation \cite{BenAchour:2020wiw}.
In this context, stealth solutions are particularly valuable, as they correspond to regular vacuum geometries of GR and thus provide ideal seeds for generating well-behaved, backreacting spacetimes. Historically, conformal transformations, a subset of disformal transformations, have been used to construct nontrivial black hole and wormhole solutions, especially in theories with conformally coupled scalar fields \cite{BBM,Bekenstein:1975ts,Martinez:2002ru,Barrientos:2022avi,Cisterna:2023uqf,Anabalon:2012tu}. In generalized scalar–tensor theories, this role is now fulfilled by disformal transformations (see \cite{BenAchour:2024hbg} for a review).

Apart from solutions in minimally coupled Einstein–scalar models, most known black holes with scalar hair are static and spherically symmetric, except for two notable stealth families \cite{Charmousis:2019vnf,Babichev:2017guv}. This limitation reflects the complexity of the field equations and the current lack of advanced solution-generating techniques. These two exceptions describing rotating stealth black holes are particularly significant.
For stealth configurations, the difficulty does not lie in solving for the metric, which remains the Kerr vacuum geometry of GR, but in identifying the scalar–tensor theory that admits the stealth solution and determining the corresponding scalar field profile.
These geometries form the focus of the present work. To date, only one fully backreacting rotating black hole, known as the disformal Kerr solution, has been constructed \cite{BenAchour:2020fgy, Anson:2020trg}. This solution arises from a disformal transformation applied to a Kerr stealth black hole within a specific DHOST theory, featuring a scalar field with both radial dependence and linear time evolution \cite{Charmousis:2019vnf}. As will be discussed, the resulting spacetime is non-circular, making its line element particularly intricate and complicating the identification of Killing horizons, ergospheres, and other intrinsic properties of rotating black holes.

In this work, we present a second fully backreacting rotating black hole with scalar hair, notable for preserving circularity. This solution is generated from a different seed—a Kerr stealth black hole within Horndeski theory—whose scalar field respects the same symmetries as the metric, namely stationarity and axisymmetry \cite{Babichev:2017guv}. By applying the same class of disformal transformations as before, and using a symmetry-compatible scalar profile, no convective effects arise, ensuring that circularity is maintained. As a result, Killing horizons, ergospheres, and other key geometric properties can be readily determined. This new exact solution thus provides an improved version of the disformal Kerr black hole which allows to closely mimic the properties of the Kerr black hole while still introducing non-trivial deviations that can be confronted to current observations. Moreover, following \cite{BenAchour:2024hbg}, we discuss how appropriate non-shift symmetric disformal transformation can be used to build asymptotically stealth Kerr black holes. 

This work is organized as follows. In Sec.~\ref{sec:prelim}, we review the necessary preliminaries, including the Kerr line element and the specific Kerr stealth solution used in our construction. In Sec.~\ref{sec:kerr-disf}, we present the new configuration, termed the Circular Disformal Kerr, and examine its main geometric properties such as horizons, singularities, special limits, and algebraic classification in Sec.~\ref{sec:main-properties}. We finish with concluding remarks in Sec.~\ref{sec:remarks}.

\section{Preliminaries}
\label{sec:prelim}

To keep the discussion as self-contained as possible, we first present two essential preliminaries before introducing our new solution.
First, since our construction is based on the Kerr metric, we briefly review the key features of Kerr spacetime, including the structure of its horizons, the formation of the ergosphere, and the nature of its singularity.
Second, we revisit the seed configuration—a Kerr stealth black hole within a specific subsector of Horndeski theory. This review identifies the class of theories that admit the seed solution and specifies the corresponding scalar field profile, which is crucial for preventing convective effects that would otherwise break circularity.
With these foundations, the following section introduces our novel geometry and examines its main properties.

\subsection{Short review of the Kerr geometry}

The Kerr line element was originally expressed in retarded Eddington–Finkelstein coordinates, which facilitate a partial extension of the geometry across the event horizon. However, to analyze the locations of horizons and ergospheres more effectively, it is convenient to use Boyer–Lindquist coordinates. These coordinates simplify the identification of such surfaces and reduce the number of off-diagonal metric components, making the metric’s circularity explicit. The Boyer–Lindquist system is obtained from the retarded null coordinates by introducing the asymptotic time coordinate. The resulting loss of the Kerr–Schild form is immaterial for the purposes of this analysis. Hence, we consider the line element in Boyer–Lindquist coordinates $(t,r, \theta, \phi)$, which takes the form
\begin{widetext}
\begin{align}
\dd{s^2_\mathrm{Kerr}} & = - \left( 1 - \frac{2m r}{\Sigma^2}\right) \dd{t}^2 - \frac{4 a m r \sin^2{\theta}}{\Sigma^2} \dd{t}\dd{\phi} + \Sigma^2 \left( \frac{\dd{r}^2}{\Delta} + d{\theta}^2 \right) + \left( r^2 + a^2 + \frac{2a^2 m r \sin^2{\theta}}{\Sigma^2}\right)\sin^2\theta \dd{\phi}^2, \label{BLKerr}
\end{align}
\end{widetext}
where we have defined the functions $\Sigma^2$ and $\Delta$ via the standard expressions
\begin{align}
\Sigma^2 = r^2 + a^2 \cos^2{\theta}, \quad \Delta = r^2 - 2M r + a^2,
\end{align}
with the parameters $m$ and $a$ representing, as usual, the mass and angular momentum per unit mass parameters. 

The Kerr geometry is a member of the Weyl–Lewis-Papapetrou (WLP) class of spacetimes and, as such, exhibits invariance under the action of the symmetry group $\mathbb{R} \times SO(2)$. This symmetry is generated by a pair of commuting Killing vector fields, $\partial_t = \xi^\mu \partial_\mu$ and $\partial_\phi = \chi^\mu \partial_\mu$, which will hereafter be collectively denoted by $(\xi, \chi)$. 

According to Papapetrou’s theorem, in vacuum the orbits of the symmetry group $\mathbb{R} \times SO(2)$—the transitivity surfaces—are everywhere orthogonal to a family of hypersurfaces known as meridional surfaces, which are defined by the non-Killing coordinates $(r, \theta)$. Consequently, the Kerr metric is said to be circular. From a geometrical standpoint, circularity can be interpreted as the absence of convective (or meridional) flows within the spacetime.
Hence, the two-planes orthogonal to the Killing vector fields $(\xi, \chi)$ are integrable,
\begin{equation}
\xi_{[\mu}\chi_\nu \nabla_\rho \xi_{\sigma]} = 0, \quad \xi_{[\mu}\chi_\nu \nabla_\rho \chi_{\sigma]} = 0, \label{frob}
\end{equation}
implying that the metric can be written in a block-diagonal form containing only a single off-diagonal term, $\dd{t}\dd{\phi}$.

For a stationary, asymptotically flat vacuum spacetime possessing a non-degenerate event horizon, the rigidity theorem applies. According to this result, stationarity necessarily implies axisymmetry, and consequently, the event horizon is a rigid Killing horizon generated by the Killing vector field $\eta = \xi + \Omega_H \chi$, which undergoes rigid rotation with a constant angular velocity $\Omega_H$.

This is the case for the Kerr geometry, where the Killing field $\eta^\pm = \xi + \Omega_H^\pm \chi$ generates two distinct Killing horizons, located at 
\begin{equation}
r_\pm=m\pm\sqrt{m^2-a^2}, \label{KerrHorizon}
\end{equation}
which correspond to the zeros of the metric function $\Delta$ and that rotate with angular velocity 
\begin{equation}
  \Omega_{\pm} = \frac{a}{r_{\pm}^2 + a^2}. \label{rotkerr}
\end{equation}
Here, $r_+$ denotes the outer (event) horizon, while $r_-$ represents the inner (Cauchy) horizon.

In this context, circularity plays a fundamentally important role. If a spacetime is non-circular (i.e., its line element is not block-diagonal), the Killing vector fields $\xi$ and $\chi$ fail to define integrable two-surfaces and are therefore twisted. As a result, a linear combination of the form $\eta = \xi + \Omega(x^i) \chi$ cannot, in general, be tangent to null generators that are hypersurface-orthogonal to a putative horizon.
This breakdown of integrability implies that the rigidity theorem does not necessarily hold, and thus the existence of a Killing horizon is no longer guaranteed. Consequently, a non-circular spacetime may lack a Killing horizon even if it remains stationary and axisymmetric. Ultimately, the failure of circularity leads to differential rotation along the horizon, preventing the existence of a single, globally defined Killing generator.

This lack of circularity introduces significant complications, preventing newly proposed rotating geometries from being examined exhaustively from a geometric standpoint. Moreover, it may raise concerns regarding the theoretical robustness and internal consistency of such a spacetime.
However, it is important to recall that Papapetrou’s theorem can still be satisfied for non-vacuum configurations. In such cases, matter fields may be incorporated without necessarily introducing convective effects, if constrained properly. This is precisely the approach that will be adopted in the following section, where our novel spacetime is introduced.

For completeness, let us briefly recall some of the remaining and most relevant geometric features of the Kerr spacetime. The loci at which the asymptotically timelike Killing vector field $\xi$ becomes null define the so-called ergospheres,
\begin{equation}
\xi^{\mu} \xi_{\mu}\big|_{r_E^\pm} = 0 \rightarrow r_E^{\pm} = m \pm \sqrt{m^2 - a^2 \cos^2{\theta}}.
\end{equation}
These surfaces delimit the regions where the frame-dragging effect induced by the rotating source becomes so strong that no observer can remain stationary with respect to infinity. Within the ergoregion, all particles and fields are compelled to co-rotate with the black hole, and static observers can no longer exist. These regions are timelike and therefore do not constitute causal boundaries; consequently, they can be traversed in both directions. Moreover, the region inside the ergosphere allows for negative energy states for neutral particles, a property that is central to extracting spinning energy from the rotating hole.

The singularity structure of the Kerr geometry differs profoundly from that of its static counterpart. A direct analysis of the Kretschmann invariant reveals the existence of a curvature singularity at
\begin{equation}
\Sigma^2 = r^2 + a^2 \cos^2\theta = 0,
\end{equation}
which occurs for $r = 0$ and $\cos\theta = 0$, the latter corresponding to the equatorial plane. This singularity is not point-like; rather, only trajectories approaching the region $r = 0$ along the equatorial plane encounter it. Paths approaching $r = 0$ from any other direction reach regular regions of the spacetime.
Indeed, the surfaces defined by $r = 0$ and $t = \text{constant}$ form a disc of radius $a$ embedded in a locally flat two-dimensional space. The singularity itself lies on this disc at $\theta = \pi/2$, thereby exhibiting a ring-like structure.

Lastly, it is worth recalling that the Kerr geometry is algebraically special within the Petrov classification, specifically of type D. Its Weyl tensor admits two distinct principal null directions and, consequently, the Kerr solution can be regarded as a particular member of the broader class of expanding Plebański–Demiański geometries \cite{Stephani:2003tm}.

\subsection{The Kerr stealth seed}\label{seed}

As discussed above, stealth black hole configurations provide excellent seed candidates for generating well-behaved, backreacting solutions in generalized scalar–tensor theories through disformal mappings, which, in their most general form, are defined as
\begin{equation}
    g_{\mu\nu}\rightarrow C(\varphi,X)g_{\mu\nu}+D(\varphi,X)\partial_\mu\varphi\partial_\nu\varphi, \label{disformalmap}
\end{equation}
where the conformal and disformal functions, $C(\phi, X)$ and $D(\phi, X)$, depend on the scalar field and its kinetic term $X=-\frac{1}{2} g^{\mu\nu} \partial_\mu \varphi  \partial_\nu \varphi$.
Generalized scalar–tensor theories are interconnected through such mappings. When the transformation depends only on the scalar field, Horndeski theory—which yields second-order field equations—remains closed under the mapping \cite{Langlois:2018dxi}. However, dependence on $X$ naturally extends the theory beyond Horndeski, into its higher-order generalizations such as DHOST theories.

The most relevant case to explore is the one in which rotation appears. In this context, two Kerr stealth black hole configurations have been identified, the best studied of which arises within a well-motivated DHOST model where gravitational waves propagate at the speed of light \cite{Babichev:2017guv}.

In this solution, the regularity of the scalar field relies on a linear time dependence that breaks the stationarity of the scalar itself while preserving the stationarity of the Einstein equations due to its linear nature. This feature is crucial: together with a radial (and, when a cosmological constant is present, possibly polar) dependence of the scalar profile, it ensures that the kinetic term remains constant, $X = X_0$. This condition greatly simplifies the associated energy–momentum tensor, allowing for a direct identification of the underlying theory and facilitating the integration of the scalar field profile. This condition has been shown to coincide with the Hamilton–Jacobi equation, implying that the scalar field profile corresponds directly to the Hamilton–Jacobi potential governing geodesic motion in the Kerr spacetime.

The disformally transformed Kerr spacetime obtained from this seed naturally becomes non-circular. The linear time dependence of the scalar field mixes with its radial component, producing a convective effect that breaks circularity. As a result, a cross-term of the form $g_{tr}$ appears, preventing the metric from being expressed in a block-diagonal form \cite{Anson:2020trg}. 

In light of the previous discussion on the Kerr geometry, the lack of circularity is problematic. The non-circular disformal Kerr spacetime possesses no Killing horizons, as shown in \cite{BenAchour:2020fgy}, since any linear combination of the form $\eta = \xi + \Omega(x^i)\chi$ necessarily requires the function $\Omega$ to depend on the polar coordinate.
Consequently, the identification of an event horizon becomes highly nontrivial. A tentative null hypersurface fulfilling this role has been proposed, but no conclusive results have yet been established.

At this stage, the use of the second Kerr stealth configuration available in the literature becomes particularly relevant. In \cite{Babichev:2017guv}, a class of Horndeski theories was identified that admits a Kerr stealth black hole solution, provided the theory is restricted to its purely quadratic sector.
Two of the assumptions underlying the no-hair theorem for Horndeski theories are violated in this particular case \cite{Capuano:2023yyh}. Specifically, the theory under consideration does not include a standard kinetic term; and, as it is obvious, the scalar field does not vanish asymptotically, but instead, its gradient tends to a constant value at spatial infinity. For clarity, the action describing the theory that supports our seed configuration is given by
\begin{equation}
S=\int \sqrt{-g} \dd[4]{x} \left\{ G_4(X) R + G_{4X} \left[ (\square \varphi)^2 - (\nabla_\mu \nabla_\nu \varphi)^2 \right] \right\},
\end{equation}
where the function $G_4(X)$ is subject to the conditions $G_{4,X}(X_0)=0$ and $G_{4,XX}(X_0)=0$. Consequently, any theory of the form
\begin{equation}
G_4(X) = \kappa + \sum_{i \geq 2} \beta_i (X - X_0)^i,
\end{equation}
admits a GR vacuum solution with a non-trivial scalar field profile that can be integrated from the condition $X=X_0$.
In particular, when considering the Kerr metric expressed in Boyer–Lindquist coordinates \eqref{BLKerr}, the scalar configuration—or scalar dress—associated with the Kerr stealth solution takes the form
\begin{equation}
\varphi(r,\theta) = \sqrt{-2X_0} \left[ a \sin\theta - \sqrt{\Delta} - m \ln(  r - m +\sqrt{\Delta}) \right], \label{seedprofile}
\end{equation}
a relation that explicitly demonstrates the regularity of the scalar field throughout the entire spacetime region, from the event horizon 
$r_+$ to spatial infinity.

The principal advantage of employing this solution as a seed lies in the fact that the scalar field profile respects the symmetries of the Kerr line element. Consequently, no mixing occurs between the transitivity and meridional coordinates, ensuring the absence of any convective effects that could otherwise break the circularity of the disformally transformed solution. In the following section, we shall demonstrate this property explicitly through the direct construction of the resulting geometry.

\vspace{0.5cm}

\section{Construction of a new Disformal Kerr spacetime}
\label{sec:kerr-disf}

As previously discussed, a primary goal of contemporary research in modified gravity is the construction of analytic exact solutions that describe rotating black holes. Of particular interest are configurations that closely emulate the Kerr geometry. This requirement entails not only that the metric deformations be controlled by a small, dimensionless parameter—so that they may be treated as perturbative corrections—but, more importantly, that if a non-perturbative regime exists, most of the fundamental properties of the Kerr solution are inherited. Among these essential features are the existence of well-defined Killing horizons, the separability of the wave equation, the Petrov type D algebraic classification, and the circularity of the spacetime geometry.

In line with the use of disformal transformations as a primitive solution-generating technique, we present here a novel disformally transformed Kerr black hole. The construction is based on the Kerr stealth solution discussed in the previous section. To apply the disformal transformation \eqref{disformalmap}, we first compute the gradient of the scalar field profile given in \eqref{seedprofile}, which yields
\begin{equation}
\partial_r \varphi = -r \sqrt{\frac{-2X_0}{\Delta}} , \quad
\partial_{\theta} \varphi = \sqrt{-2X_0} a \cos{\theta}.
\end{equation}

As a first step in our analysis, we consider the conformal and disformal factors to take the simple forms $C(\varphi,X)=C_0$ and  $D(\varphi,X)=D_0$. The specific choice of the conformal factor is not of primary importance for our present purposes, as it is typically introduced to establish a correspondence between two particular theories, often in conjunction with a redefinition of the scalar field. The essential ingredient in our construction is the disformal factor. Although, in principle, it can be taken as a general function of $\varphi$ and $X$, for simplicity, we restrict our analysis to the case of a constant disformal factor, in accordance with previous studies \cite{Anson:2020trg}. Nevertheless, toward the end of this work, we shall consider a more general functional form for the disformal factor, which will always be of the form $D=D(\varphi)$, so that we remain in a Horndeski model.

With these considerations in place, the disformally transformed Kerr solution, expressed as an exact modification of the Kerr line element \eqref{BLKerr}, can be written in the following form:
\begin{widetext}
    \begin{equation}
    \dd{s^2}=C_0 \dd{s^2_\mathrm{Kerr}}- D_0 \left\{  \frac{2X_0r^2}{ \Delta}  \dd{r}^2  - \frac{4X_0}{\sqrt{\Delta} } ar \cos{\theta} \dd{r} \dd{\theta} + 2X_0 a^2 \cos^2{\theta} \dd{\theta}^2  \right\}.
\end{equation}
\end{widetext}
As expected, owing to the scalar field profile being compatible with the underlying symmetries, the structure of the metric within the transitivity submanifold 
$(t,\phi)$ remains unaltered. Consequently, the resulting geometry preserves the initial circularity of the seed.

It is straightforward to verify that the newly obtained spacetime is not Ricci flat, $R\neq0$. This outcome is intuitively expected, as there is no compelling reason for the disformal mapping to preserve the stealth character of the seed solution—namely, the property that the field equations are satisfied in such a way that 
$G_{\mu\nu}=0=T_{\mu\nu}$\footnote{Such stealth configurations were first identified in \cite{Henneaux:2002wm} for $2+1$ Einstein-scalar theory.}, for the non-trivial scalar field\footnote{A notable exception appears in five dimensions in the case of a rotating Myers–Perry black hole with two equal angular momenta \cite{Baake:2021kyg}.}.

The resulting solution thus represents a line element with a non-trivial backreaction from the scalar field, which satisfies the field equations of a new theory. For the purposes of our construction, it is not necessary to present this theory explicitly here. It suffices to note that, like the initial model, the target theory belongs to a quadratic Horndeski class in which a canonical kinetic term is absent.

Before proceeding with a geometrical dissection of our solution, let us first express it in a form that makes the overall impact of the disformal transformation on each of the metric components more transparent
\begin{widetext}
    \begin{align}
\dd{s^2} & = C_0 \left[ -  \left( 1 - \frac{2m r}{\Sigma^2}\right) \dd{t}^2 - \frac{4 a m r \sin^2{\theta}}{\Sigma^2} \dd{t}\dd{\phi} + \left( r^2 + a^2 + \frac{2a^2 m r \sin^2{\theta}}{\Sigma^2}\right) \sin^2{\theta} \dd{\phi}^2 \right]  \nonumber \\
& + \Sigma^2 \left[ \left( \frac{C_0}{\Delta} - \frac{2D_0 X_0 r^2}{\Delta \Sigma^2}  \right)\dd{r}^2 + \left( C_0 - \frac{2D_0X_0 a^2 \cos^2{\theta}}{\Sigma^2}  \right) \dd{\theta}^2 + \frac{4D_0 X_0}{ \sqrt{\Delta} \Sigma^2} ar \cos{\theta} \dd{r}\dd{\theta} \right]\label{disformalKerr}.
\end{align} 
\end{widetext}
In this representation, the full modification of the meridional sector of the metric becomes apparent, particularly the emergence of a non-diagonal term. However, this term can be eliminated through an appropriate diffeomorphism, albeit at the cost of introducing more complicated functional forms for the remaining metric components. However, notice that its asymptotic form at large $r$ will introduce a conical defect, which will arise as an overall constant next to the $g_{rr}$ metric component. As we shall discuss below, this defect can be removed by upgrading the disformal factor to a suitable function of $\varphi$.  Finally, since it will be useful in what follows, we provide the expression for the inverse metric
\begin{widetext}
\begin{align}
    g^{\mu\nu} \partial_\mu \partial_\nu =& \frac{1}{C_0 \Delta \Sigma^2} \Bigg\{- \big[a^2 (r^2 + a^2) \cos^2\theta + r(r^3 + a^2 r + 2 a^2 M \sin^2\theta)\big]\partial_t^2 - 4 a M r \partial_t\partial_\phi \nonumber\\
    &+ \frac{\Delta^2}{\Sigma^2}\frac{C_0 r^2 + (C_0 -2 D_0 X_0) a^2 \cos^2\theta}{C_0 -2 D_0 X_0} \partial_r^2 - \frac{\Delta^{3/2}}{\Sigma^2}\frac{4a D_0 X_0 r \cos\theta}{C_0 -2 D_0 X_0} \partial_r \partial_\theta \nonumber\\
    &+ \frac{\Delta}{\Sigma^2}\frac{r^2 (C_0 -2 D_0 X_0) + a^2 C_0 \cos^2\theta}{C_0 -2 D_0 X_0} \partial_\theta^2 + \frac{1}{\sin^2\theta}\big[r(r-2M) + a^2 \cos^2\theta\big] \partial_\phi^2 \Bigg\} \,, 
    \label{eq:inverse-metric}
\end{align}
\end{widetext}
where $C_0-2D_0X_0\neq0.$

Ergoregions of a spacetime are the surfaces beyond which no observer can remain stationary. It can be readily observed that these surfaces are unchanged relative to those of the Kerr seed. Indeed, the Killing vector $\xi$ retains its norm; consequently, the locations of the ergospheres remain unaltered. Therefore, the ergospheres of our solution are bounded, and the Killing vector $\xi$ remains timelike throughout the entire asymptotic region.

Again, since the $(t-\phi)$-sector of the metric remains unaltered, the component $g_{\phi\phi}$ preserves its original form; consequently, no chronology horizon appears within the domain of outer communications (as defined below in terms of the Killing horizons). The solution thus remains chronal, that is, free of closed timelike curves in the same region as the Kerr spacetime.

\section{Main properties of the solution}
\label{sec:main-properties}

In this section, we investigate the main properties of the spacetime~\eqref{disformalKerr}, namely the position of the event horizon, the Petrov type and the large $r$ limit. In particular, we shall see that investigating the position of the event horizon is made easy thanks to the preserved circularity, which ensures the existence of a Killing horizon.

\subsection{Killing horizon}
\label{sec:horizon}

Owing to the circularity of the spacetime~\eqref{disformalKerr}, we expect the existence of a Killing horizon and, consequently, that the horizon of the circular disformal Kerr black hole rotates rigidly with a constant angular velocity $\Omega_H$. In this section, we determine the location of the Killing horizon. First, let consider the vector field $\eta^\mu$ defined by
\begin{equation}
    \eta^\mu \partial_\mu = \partial_t + \Omega_H \partial_\phi \,,
\end{equation}
with $\Omega_H$ as given by \eqref{rotkerr} with the plus branch. 
%
%
This vector is a linear combination of the vectors $\partial_t$ and $\partial_\phi$. Since the metric~\eqref{disformalKerr} does not depend on $t$ nor on $\phi$, these two vectors are Killing vectors and $\eta^\mu$ is one as well. Furthermore, in the original Kerr geometry, it corresponds to a Killing (outer event) horizon located at $r = r_+ = M + \sqrt{M^2 - a^2}$. As we will see, this is still the case in the disformed geometry.

Let us then look for null hypersurfaces of the geometry~\eqref{disformalKerr}. We look for surfaces parametrized by $r = f(\theta)$. Such an hypersurface is null if and only if
\begin{equation}
    \left.g^{\mu\nu} \partial_\mu\big[r - f(\theta)\big] \partial_\nu\big[r - f(\theta)\big]\right|_{r=f(\theta)} = 0 \,.
\end{equation}
This implies
\begin{equation}
    g^{rr} - 2 g^{r\theta} f'(\theta) + g^{\theta\theta} f'(\theta)^2 = 0 \,.
\end{equation}
From~\eqref{eq:inverse-metric}, we see that the component $g^{rr}$ of the inverse metric is proportional to $\Delta$, which implies that the hypersurface $\mathcal{H}$ defined by $r = r_+$ is a null hypersurface. Furthermore, the norm $\eta^\mu \eta_\mu$ of the Killing vector is zero on $\mathcal{H}$, and we can check explicitly that $\eta^\mu$ is orthogonal to $\mathcal{H}$. These three conditions imply that $\mathcal{H}$ is a Killing horizon of the disformed spacetime.

As a null hypersurface, $\mathcal{H}$ is a one-way membrane: any particle that crosses it from the black hole exterior can never exit it again. To prove that it is a black hole event horizon, we further need to find at least one geodesic that starts away from $\mathcal{H}$ and escapes to null infinity. This implies that there is no hypersurface on the exterior of $\mathcal{H}$ that has the properties of an event horizon; therefore, $\mathcal{H}$ is the outer event horizon.

\begin{figure}
    \centering
    \includegraphics{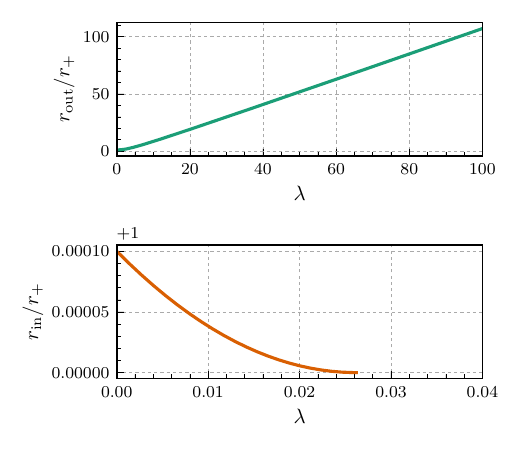}
    \caption{Evolution of $r(\lambda)$ for the two geodesics $\mathcal{L}_\mathrm{out}$ and $\mathcal{L}_\mathrm{in}$. We see that $\mathcal{L}_\mathrm{out}$ evolves towards infinity while $\mathcal{L}_\mathrm{in}$ falls through $\mathcal{H}$. We have taken $M = 1$, $a = 0.95$, $C_0 = 1$, $B_0 = 10$, $X_0 = 1$ and $\epsilon = 10^{-4}$.}
    \label{fig:geodesics}
\end{figure}

We obtain such a geodesic numerically. Let us consider two geodesics $\mathcal{L}_\mathrm{out}$ and $\mathcal{L}_\mathrm{in}$ with initial tangent vectors $p^\mu \partial_\mu = \pm \partial_r$ and affine parameter $\lambda$, passing through the point $(t=0, r = r_+(1+\epsilon), \theta = \pi/4, \phi = 0)$ at $\lambda=0$. We integrate the geodesics equation and plot the results in Fig~\ref{fig:geodesics}. We observe that $\mathcal{L}_\mathrm{out}$ does evolve towards null infinity, while $\mathcal{L}_\mathrm{in}$ falls into $\mathcal{H}$. Following the reasoning detailed above, this implies that $\mathcal{H}$ is the event horizon of the disformed spacetime.

Now, it is also possible to prove that $\mathcal{H}$ is the event horizon of our geometry after diagonalizing it. 
Any two-dimensional Riemannian geometry is, at least locally, conformally flat. Consequently, the meridional submanifold --- whose metric can generically be expressed as $g_{rr}\dd{r}^2+2g_{\theta r}\dd{\theta}\dd{r}+g_{\theta\theta}\dd{\theta}^2$ --- may always be recast via the introduction of isothermal coordinates $(u,v)$ into the diagonal form
\begin{equation}
\dd{s}_\mathrm{meridional}^2 = \Omega(u,v)(\dd{u}^2 + \dd{v}^2),
\end{equation}
for some $\Omega>0$. Achieving this representation requires solving the first-order elliptic Beltrami equation, for which exact solutions exist only in very special and tractable cases. Nevertheless, in our case, the condition $g_{uu}=g_{vv}$ is not essential, since we will not employ a WLP form for our final line element. However, diagonalizing the meridional submanifold alone is conveninent for the computation of several spacetime observables. While we will not need the diagonal form for the computations presented in this paper, we detail the diagonalization for completeness and future computations.

In order to perform this diagonalization, we define a new coordinate $R(r, \theta)$. The metric on the new meridional submanifold defined by the coordinates $(R, \theta)$ is diagonal if $r$ takes the form
\begin{equation}
    r = R + f(u) \,,\quad u = \artanh\Big(\frac{a \sin\theta}{\sqrt{R^2 + a^2}}\Big) \,,
\end{equation}
where $f$ is solution to the following differential equation
\begin{widetext}
    \begin{equation}
        -2D_0 X_0 f \sqrt{f- R_+} \sqrt{f - R_-} + \dv{f}{u} \frac{\cosh(u)^2}{\sqrt{R^2 + R_+ R_-}} \Big[C_0 (R^2 + R_+ R_-) \tanh(u)^2 - (C_0 -2 D_0 X_0) f^2 - C_0 R_+ R_-\Big] = 0 \,.
        \label{eq:f-chgmt-coords}
    \end{equation}
\end{widetext}
To avoid confusion with the old radial coordinate $r$, we have defined $R_{\pm}=m\pm\sqrt{m^2-a^2}$ to denote the horizon loci. 

Equation \eqref{eq:f-chgmt-coords} admits no apparent closed-form solution, which is why we solve it perturbatively in $D_0$.
This approach allows us to diagonalize the metric and explicitly determine the location of the event horizon.
We have carried out the computation up to order five in $D_0$, but for clarity we present only the first- and second-order results here.
The change of coordinates reads 
\begin{widetext}
    \begin{align}
& r  = R - \frac{2D_0 X_0}{C_0} \frac{\sqrt{(R-R_{+})(R-R_{-}) }}{\sqrt{R^2+ R_{+} R_{-}}} u  \nonumber\\
& -  \frac{4D^2_0  X^2_0 R}{4C^2_0(R^2+ R_{+} R_{-})^2 } \left[ u \left[ \left[ u (  R^3 (R_{+} + R_{-})  -2 (R^4 + R^2_{+} R^2_{-}) + 3 R_{-} R_{+} (R_{+} + R_{-}) R \right] \right. \right.\nonumber\\
&\;  \left. \left. - 2 \left(\sqrt{(R-R_{+})(R-R_{-}) }\sqrt{R^2- R_{+} R_{-}} - 2 R_{+} R_{-} u) \right] - R^2(R-R_{+})(R-R_{-}\right) \cosh{(2u)} \right.\nonumber \\
& \; \left.+ R^2 \sinh{(2u)} \left( \sqrt{(R-R_{+})(R-R_{-}) } \sqrt{R^2- R_{+} R_{-}} + 2 R^2 u + 2R_{-} R_{+} u - 2 R (R_{+} + R_{-}) u  \right) \right]  + \mathcal{O}(D^3_0) \,.
\end{align}
\end{widetext}
This procedure eliminates the off-diagonal term $g_{r\theta}=0$ without introducing any additional off-diagonal components, although it renders the metric algebraically more involved at second order. It is observed that the metric component $g^{RR}$ up to fifth order 
\begin{equation}
    g^{RR}=g^{RR}_0+D_0g^{RR}_1+D_0^2g^{RR}_2+D^3_0g^{RR}_3+D_0^4g^{RR}_4+\mathcal{O}(D_0^5),
    \end{equation}
remains proportional to the function $\Delta$.
\subsection{Slowly rotating and large $r$ limits}

Now that we have obtained a form of the metric free of the off-diagonal term, we examine its large-$r$ behavior and the slow-rotation limit of the geometry.
Let us first consider the large-$r$ expansion of the new solution, which at leading order takes the form
\begin{widetext}
\begin{align}
\dd{s}^2  = C_0\Biggl[-   \left( 1- \frac{2M}{R}\right)  \dd{t}^2 - \frac{4 a M }{R} \sin^2{\theta} \dd{t}\dd{\varphi}\Biggr.   
  +  \left(1 - \frac{2D_0 X_0}{C_0}\right) \left( 1+ \frac{2M}{R}\right) \dd{R}^2 +  \rho^2 \dd{\theta}^2\nonumber
\Biggl.+ R^2\sin^2{\theta}\dd{\phi}^2\Biggr] \,.
\end{align}
\end{widetext}
In this form, the conical defect becomes manifest. This is not problematic, however, as it can be removed by an appropriate disformal factor --- for example, one of the form $D(\varphi)=D_0/\varphi$, as shown in Sec.~\ref{sec:curing-conical}. Thus, for the purposes of a purely geometric analysis of our solution, this issue is not immediately relevant.

On the other hand, the slowly rotating limit of the solution takes the form
\begin{widetext}
   \begin{align}
   \dd{s}^2 & = - C_0\left[  \left( 1 - \frac{2M}{R}\right) + \frac{4 D_0 X_0 }{(C_0 -2 D_0 X_0)} \sqrt{\frac{R-2M}{R^5}} aM  \sin{\theta} \right] \dd{t}^2  - \frac{4 aM C_0 \sin^2{\theta}}{R} \dd{t}\dd{\phi} \nonumber \\
& +  C_0 R^2 \sin^2{\theta} \left( 1 - \frac{4 D_0 X_0}{C_0 -2 D_0 X_0} \sqrt{R(R-2M)} a \sin{\theta} \right) \dd{\phi}^2 \nonumber \\
& + C_0\left(1 - \frac{2D_0 X_0}{C_0}\right) \left( 1 - \frac{2M}{R}\right)^{-1} \dd{R}^2 +\ C_0 \left[ 1 - \frac{4 a D_0 X_0}{C_0 -2 D_0 X_0} \frac{\sqrt{R(R-2M)}}{R^2} \sin{\theta}\right] R^2 \dd{\theta}^2. \label{slowlydisformal}
\end{align}
\end{widetext}
In this limit, the horizon, located at the pole of $g_{RR}$, lies at $R = 2M$, as expected. The location of the ergosphere can be determined by solving $g_{tt} = 0$. At first order in $a$, this equation again yields $R = 2M$, showing that the ergoregion is not modified at linear order in the spin, in agreement with the GR result for the Kerr solution. At second order in $a$, the ergosphere becomes deformed, with the deformation parametrized by the value of $D_0$. In the static limit $a=0$, one simply recovers a metric close to Schwarzschild, supplemented by an additional conical defect.
This slowly rotating limit is useful because it helps clarify the form of the fully rotating metric (a cumbersome expression we do not write explicitly here) once the meridional submanifold has been diagonalized perturbatively in $D_0$.

\subsection{Petrov type}

Given a newly obtained geometry, it is customary to determine its algebraic type, which, as is well known, can be identified through the Petrov classification \cite{Stephani:2003tm}. This analysis serves several important purposes: it helps assess the genuine novelty of the geometry (ensuring that it is not merely a reparameterization of a known solution) and provides insights into various structural and dynamical properties, such as kinematic integrability, the presence of hidden symmetries, and the separability of the geodesic equations, among others. Although the Petrov classification can, in principle, be carried out with respect to an arbitrary null tetrad, in this work we begin with an aligned null tetrad—namely, one that is aligned with the principal null directions of the spacetime. Based on the known behavior of disformal transformations and their effect on the algebraic type of a given seed metric \cite{Achour:2021pla}, we expect the resulting spacetime to be algebraically general, i.e., of Petrov type I. Consequently, the chosen null tetrad can, at most, be aligned with one of the four distinct principal null directions of the Weyl tensor. We therefore consider the following null tetrad 
\begin{widetext}
\begin{subequations}
\begin{align}
\ell^{\mu} \partial_{\mu} & = \frac{1}{\sqrt{C_0}} \left[ \frac{r^2+a^2}{\Delta} \partial_t +  \left( 1- \frac{\alpha r^2 \sqrt{-2X_0}}{\Sigma^2}\right) \partial_r + \frac{a \sqrt{-2X_0} \alpha r \cos{\theta}}{\Sigma^2 \sqrt{\Delta}}\partial_{\theta} + \frac{a}{\Delta} \partial_{\phi}
\right] \\
n^{\mu} \partial_{\mu} & = \frac{1}{\sqrt{C_0}} \left[  \frac{r^2+a^2}{2 \Sigma^2} \partial_t - \frac{\Delta}{2\Sigma^2} \left( 1 -\frac{\alpha r^2 \sqrt{-2X_0}}{\Sigma^2}  \right) \partial_r - \frac{a \sqrt{-2X_0} \alpha r \cos{\theta} \sqrt{\Delta}}{\Sigma^2 }\partial_{\theta} + \frac{a}{2 \Sigma^2} \partial_{\phi}
 \right] \\
m^{\mu} \partial_{\mu} & = \frac{1}{\sqrt{2C_0}} \left[ \frac{ia \sin{\theta}}{r+ ia \cos{\theta}} \partial_t + \frac{a r \sqrt{-2X_0 \Delta} \alpha \cos{\theta}}{\Sigma^2(r+ i a \cos{\theta})} \partial_r - \frac{\Sigma^2 + a^2 \sqrt{-2X_0} \alpha \cos^2{\theta} }{\Sigma^2 (r+ i a \cos{\theta})} \partial_{\theta} + \frac{i \text{cosec}{\theta}}{r + i a \cos{\theta}} \partial_{\phi} 
\right],
\end{align}
\end{subequations}
\end{widetext} 
satisfying by definition $\ell^{\mu} n_{\mu} =-1$, $m^{\mu} \bar{m}_{\mu} =+1$, and where we have defined 
\begin{equation}
\alpha=\frac{C_0-2D_0X_0-\sqrt{C_0(C_0-2D_0X_0)}}{\sqrt{-2X_0}(C_0-2D_0X_0)}.
\end{equation}
This tetrad naturally connects to the Kinnersley tetrad of the Kerr geometry in the limit $X_0 \rightarrow 0$. In this limit, the tetrad becomes fully aligned with the two repeated principal null directions characteristic of Kerr spacetime, consistent with its Petrov type D nature. In this case, the seed spacetime is algebraically characterized by the set of Newman–Penrose scalars $(\Psi_2 \neq 0, \Psi_0 = \Psi_1 = \Psi_3 = \Psi_4 = 0)$. The new configuration, however, is algebraically general, corresponding to Petrov type I, as the following relation between the curvature scalar constructed from the self-dual Weyl tensor is fulfilled
\begin{equation}
    I^3\neq27J^2, 
\end{equation}
with 
\begin{align}
    I&=\Psi_0\Psi_4-4\Psi_1\Psi_3+3\Psi_2^2,\\
    J&=\det\begin{pmatrix}
\Psi_4 & \Psi_3 & \Psi_2\\
\Psi_3 & \Psi_2 & \Psi_1\\
\Psi_2&\Psi_1&\Psi_0
\end{pmatrix},
\end{align}
where expressions for these scalars, written in terms of Newman-Penrose components of the Weyl tensor, need to be treated perturbatively in $D_0$ to be manageable. Even in this case, the expressions are too involved to be presented here. 

Finally, let us comment that the singularity structure of the new solution remains the same as in the Kerr geometry, as can be seen from the Kretschmann invariant, namely, it possesses the ring singularity defined by $r=0$ and $\theta=\pi/2$.

\subsection{Curing the conical defect}
\label{sec:curing-conical}

Now, as pointed out above, our geometry exhibits a conical defect. However, as proposed in \cite{BenAchour:2024hbg}, this can be easily removed by promoting the constant disformal factor to a function of the scalar field $\varphi$ such that the divergence in the scalar field profile is used to cancel the unwanted large distance behavior of the solution. This strategy is generic to stealth solution in which the scalar profile admits a growing asymptotic profile. Interestingly, using this trick allows to turn stealth solutions to \textit{asymptotically stealth} solution, where the deviations with respect to GR show up only in a restricted region. 

This can be achieved by considering a field-dependent disformal transformation of the form 
\begin{equation}
    g_{\mu\nu} \rightarrow g_{\mu\nu} + \frac{D_0}{\varphi} \partial_\mu \varphi \partial_\nu \varphi \,.
\end{equation}
The resulting line element reads 
\begin{widetext}
\begin{equation}
    \dd{s}^2 = \dd{s}^2_\text{Kerr} - \frac{D_0}{\varphi} \Big[\frac{2r^2 X_0}{\Delta} \dd{r}^2 + 2a^2 \cos^2\theta X_0 \dd{\theta}^2 - \frac{4 a r X_0 \cos\theta}{\sqrt{\Delta}} \dd{r}\dd{\theta} \Big] \,.
\end{equation}
\end{widetext}
The large $r$ limit of this line element, upon $C_0\rightarrow1$, now yields
\begin{equation}
    \dd{s}^2 \sim - \dd{t}^2 + \dd{r}^2 + r^2 \big(\dd{\theta}^2 + \sin^2\theta \dd{\theta}^2\big) \,,
\end{equation}
which implies that there is no conical defect in this solution. The reasoning applied in Sec.~\ref{sec:horizon} can be transposed to the present case without difficulty. We find that the horizon of the black hole is located ar $r = r_+$.

\section{Concluding remarks}\label{sec:remarks}

In this work, we have presented a novel circular, rotating solution within a generalized scalar–tensor model belonging to Horndeski gravity, a theory characterized by second-order field equations. Rotating solutions of this kind are notoriously rare. Unlike the only previously known exact configuration of this type \cite{BenAchour:2020fgy, Anson:2020trg}, our solution is circular. This feature is made possible by employing a Kerr stealth configuration endowed with a scalar field that shares the same symmetries as the Kerr metric—namely, stationarity and axisymmetry \cite{Babichev:2017guv}. The resulting spacetime is no longer stealth and is generically non–Ricci-flat, as expected. Several key properties of the Kerr seed are preserved: the locations of the horizons and ergospheres remain unchanged, the domain of outer communications is free of closed timelike curves, and the structure of the ring singularity is maintained. However, the solution becomes algebraically general, in accordance with the known impact of disformal transformations on the Petrov type of a seed geometry \cite{Achour:2021pla}.

Since the black hole horizon in the circular disformed Kerr is a Killing horizon, an instructive follow-up would be the computation of the surface gravity on this surface. In GR, the zeroth law states that surface gravity must be constant on the horizon~\cite{Bardeen:1973gs}. This statement has been generalized using geometrical arguments to a theory-independent one: any bifurcate Killing horizon must lead to a constant surface gravity~\cite{Kay:1988mu}. In our case, we have not proven whether or not the Killing horizon is bifurcate, but its similarity with the Kerr horizon hints in that direction. As a result, checking whether the surface gravity is constant would be an interesting consistency check, especially considering recent works on non-circular spacetimes and non-Killing horizons~\cite{Babichev:2025szb,DelPorro:2025hse}. However, we have not been able to do so, mainly because of the additional coordinate singularity added by the disformal part in the metric. In the case of Kerr, evading the coordinate singularity of Boyer-Lindquist coordinates is done by introducing advanced Kerr coordinates, which rely on the principal null directions of spacetime; in our case, spacetime being of type I, such a construction is much more involved. We leave this investigation for future work.

It is instructive to explore how stealth scalar fields of the type considered here can be used to construct new exact solutions in Horndeski gravity that extend beyond the spherically symmetric sector. Indeed, the results of \cite{Babichev:2017guv}, summarized in section \eqref{seed}, apply to a broad class of seed geometries beyond Kerr. Because the scalar field always respects the symmetries of the underlying spacetime, its backreaction affects only the meridional submanifold of the target geometry. This behavior closely parallels that found in theories with minimally coupled scalar fields. As established by the Eris–Gürses theorem \cite{Eris:1976xj}, the full backreaction of such fields is encoded in the non-Killing sector of the metric, typically at the expense of introducing curvature singularities either at the would-be horizon or at infinity, as dictated by no-hair theorems. In the present case, this pathology is avoided: the combination of stealth configurations and disformal transformations provides a mechanism for endowing spacetimes with scalar hair without generating additional curvature singularities. It would therefore be of interest to explore the application of this approach to other geometries, such as Taub–NUT or accelerating spacetimes, for which analogous constructions are currently unknown.

\section{Acknowledgments}
A.C. is supported by the FONDECYT grant 1250318. We thank Eric Gourgoulhon for insightful discussions about the properties of Killing horizons.

\bibliography{apssamp}

\end{document}